\documentclass[runningheads]{llncs}
\usepackage[T1]{fontenc}
\usepackage{graphicx}
\usepackage{xcolor} 
\usepackage{amsfonts}
\usepackage{siunitx} 
\usepackage{booktabs}
\usepackage{orcidlink}
\usepackage{tikz}
\usepackage{float}
\usepackage{placeins}
\usetikzlibrary{positioning, arrows.meta, shapes.multipart}
\usetikzlibrary{fit, positioning, shapes, backgrounds}

\newcommand{\hi}{{{H\fontsize{8}{12}\selectfont I}}}

\newcolumntype{M}[1]{>{\centering\arraybackslash}m{#1}} 

\begin{document}
\title{Exploring the Early Universe with Deep Learning}

\titlerunning{Deep Learning \& Early Universe}
%
\author{Emmanuel de Salis\inst{1}\orcidlink{0000-0002-1221-6664},
        Massimo De Santis\inst{1}\orcidlink{0009-0001-5834-3976},
        Davide Piras\inst{2}\orcidlink{0000-0002-9836-2661},
        Sambit K. Giri\inst{3}\orcidlink{0000-0002-2560-536X},
        Michele Bianco\inst{4}\orcidlink{0000-0002-6766-0017},
        Nicolas Cerardi\inst{5}\orcidlink{0009-0004-1864-512X},
        Philipp Denzel\inst{6}\orcidlink{0000-0003-0126-0659},
        Merve Selcuk-Simsek\inst{7}\orcidlink{0000-0001-6916-5899},
        Kelley M. Hess\inst{8}\orcidlink{0000-0001-9662-9089},
        M. Carmen Toribio\inst{8}\orcidlink{0000-0001-8063-2881},
        Franz Kirsten\inst{8}\orcidlink{0000-0001-6664-8668} \and
        Hatem Ghorbel\inst{1}\orcidlink{0000-0001-5501-9807}
        }
        
\authorrunning{$SE_{ar}CH$ \textit{Science Working Group}}
\institute{Haute Ecole Arc Ing\'enierie, University of Applied Sciences \& Arts Western Switzerland (HES-SO), Saint-Imier, Switzerland \and
Département de Physique Théorique and Centre Universitaire d’Informatique, Université de Genève, Genève, Switzerland \and
Nordita, KTH Royal Institute of Technology and Stockholm University, Hannes Alf\'vens v\"ag 12, SE-106 91 Stockholm, Sweden \and
Institute for Particle Physics \& Astrophysics, ETH Zurich, Wolfgang-Pauli-Str 27, 8093 Zurich, Switzerland \and
Laboratoire d’Astrophysique, Ecole Polytechnique Federale de Lausanne EPFL, Observatoire de Sauverny, Versoix 1290, Switzerland \and
Centre for Artificial Intelligence, ZHAW Zurich University of Applied Sciences, Technikumstrasse 71, 8400 Winterthur, Switzerland \and
Institute for Data Science, FHNW University of Applied Sciences \& Arts Northwestern Switzerland, Bahnhofstrasse 6, Windisch, 5210, Switzerland \and
Department of Space, Earth and Environment, Chalmers University of Technology, Onsala Space Observatory, SE-43992 Onsala, Sweden }
\maketitle 
\begin{abstract}
Hydrogen is the most abundant element in our Universe. The first generation of stars and galaxies produced photons that ionized hydrogen gas, driving a cosmological event known as the Epoch of Reionization (EoR). The upcoming Square Kilometre Array Observatory (SKAO) will map the distribution of neutral hydrogen during this era, aiding in the study of the properties of these first-generation objects. Extracting astrophysical information will be challenging, as SKAO will produce a tremendous amount of data where the hydrogen signal will be contaminated with undesired foreground contamination and instrumental systematics. To address this, we develop the latest deep learning techniques to extract information from the 2D power spectra of the hydrogen signal expected from SKAO. We apply a series of neural network models to these measurements and quantify their ability to predict the history of cosmic hydrogen reionization, which is connected to the increasing number and efficiency of early photon sources. We show that the study of the early Universe benefits from modern deep learning technology. In particular, we demonstrate that dedicated machine learning algorithms can achieve more than a $0.95$ $R^2$ score on average in recovering the reionization history. This enables accurate and precise cosmological and astrophysical inference of structure formation in the early Universe.

\end{abstract}

\setcounter{footnote}{0}
\section{Introduction}
The Epoch of Reionization (EoR) marks a pivotal yet poorly understood phase in the early Universe, occurring within the first billion years after the Big Bang—less than 10\% of its current estimated age of 13.8 billion years \cite{PlanckCollaboration2018}. During this time, ultraviolet photons from the first stars, galaxies, and quasars gradually reionized the cold, neutral hydrogen in the intergalactic medium (IGM), completing a major phase transition in the Universe’s thermal and ionization history over approximately 500 million years \cite{Furlanetto2006CosmologyUniverse}. A key probe of this process and the presence of these primordial sources is the 21-cm signal, arising from the hyperfine transition in neutral hydrogen (\hi), which emits or absorbs radiation at a rest-frame wavelength of 21-cm and frequency of 1.42 GHz \cite{Furlanetto2006CosmologyUniverse}.

To detect this faint signal, the world's largest radio telescope -- Square Kilometre Array Observatory (SKAO)\footnote{\href{www.skao.int}{www.skao.int}} -- is under-construction and aims to observe the redshifted 21-cm emission from neutral hydrogen across cosmic timescales ranging from approximately 150 million to a few billion years after the Big Bang \cite{Koopmans2015TheArray}. Due to the expansion of the Universe, the original 21-cm wavelength is stretched (redshifted), shifting the signal into lower radio frequencies over time. This effect enables three-dimensional mapping of the neutral hydrogen distribution across different cosmic epochs, a technique known as 21-cm tomography. With its unprecedented sensitivity and resolution, SKAO's low frequency component (SKA-Low) is expected to measure the 21-cm signal from the EoR \cite{Giri2018OptimalObservations}. 

Current radio experiments, such as the Low-Frequency Array (LOFAR), already generate terabytes of data in their efforts to detect the 21-cm signal \cite{Patil2017UpperLOFAR}. The SKA will take this even further, producing petabytes of data \cite{lahav2023deep}, posing significant challenges for manual analysis and interpretation. Extracting meaningful physical constraints on the early Universe from such large datasets will require automated, scalable approaches. In this work, we explore and compare several machine learning methods for analysing simulated 21-cm signals, focusing on their effectiveness in recovering key physical parameters. These developments are essential for building a robust data analysis pipeline capable of handling the enormous data volumes expected from SKA-Low.

\section{Related work}
Machine learning techniques have shown significant potential in extracting the 21-cm signal and inferring parameters of the EoR, owing to their ability to process complex, high-dimensional data. 

Convolutional neural network (CNN) architectures are particularly suited to analyse spatial patterns within tomographic maps and spectrograms \cite{hassan2019identifying,murakami2024differentiating}, and have shown effective results in closely related tasks \cite{Bianco2021segunet,bianco2024segunet}. Artificial neural networks (ANN), including multilayer perceptron (MLP)-based models, also showed notable results in similar applications \cite{jennings2020analysing}, while autoencoders, particularly variational autoencoders (VAEs), have been successfully applied to extract signal parameters with high accuracy, even under challenging conditions \cite{tripathi2024extracting}.
Other techniques to perform robust inference include simulation-based inference (SBI), which has found extensive applications across multiple disciplines, including astrophysics \cite{von2025kids}, seismology \cite{Saoulis25}, chemistry \cite{Dingeldein24}, and more. Recent studies have demonstrated that SBI is a powerful tool for extracting the 21-cm signal \cite{prelogovic2023exploring}, particularly in scenarios with intractable or non-Gaussian likelihoods. SBI leverages neural networks to approximate posterior distributions directly from simulations, bypassing the need for explicit likelihood formulations.

\begin{figure}[t]
    \centering
    \vspace{-4mm}
    \includegraphics[width=\textwidth]{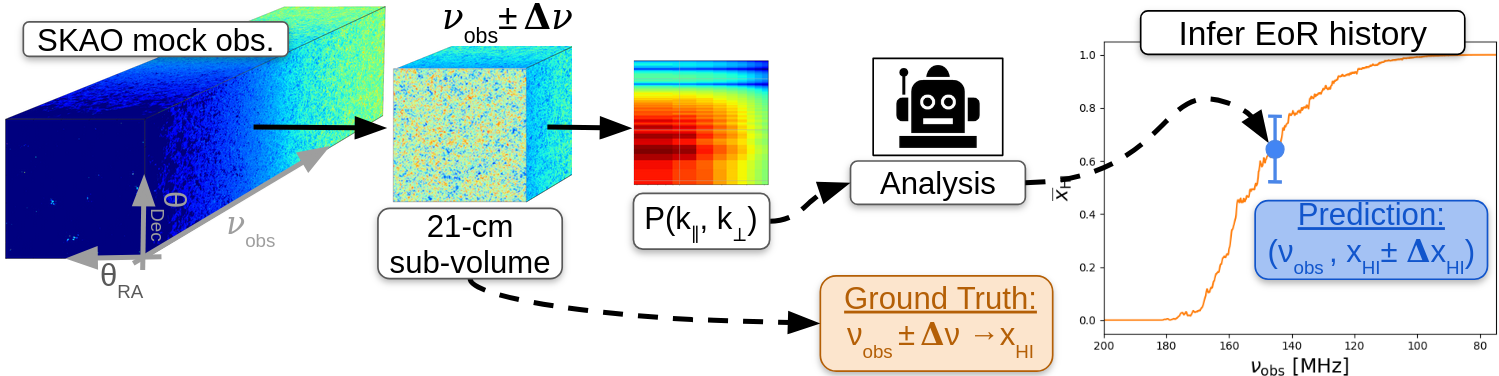}\vspace{-3mm}
	\caption{Schematic representation of our inference pipeline for one of the three frequency ranges, $\nu_\mathrm{obs}\pm \Delta\nu$ as explained in \S\ref{sec:data}.}\vspace{-4mm}
	\label{fig:pipeline}
\end{figure}

\section{Methods}
\subsection{Dataset Generation}\label{sec:data}
We produce a training set of expected data from the SKA-Low to develop machine learning methods. Radio interferometry-based telescopes, such as the SKAO, can reconstruct fluctuations in the differential brightness temperature $\delta T_\mathrm{b}$ at a given position on the sky $\mathbf{r}$ and the frequency at which it is observed $\nu_\mathrm{obs}$, thus $\delta T_\mathrm{b} (\mathbf{r}, \nu_\mathrm{obs}) \propto x_{\rm HI}(\mathbf{r}, \nu_\mathrm{obs})$ \cite{Furlanetto2006CosmologyUniverse}. This three-dimensional data is referred to as tomographic 21-cm signal data, where the values of $\nu_\mathrm{obs}$ corresponds to different cosmic time. This data is sensitive to the spatial and temporal evolution of $x_{\rm HI}$, quantifying the fraction of neutral hydrogen (\hi) in the IGM during the EoR, which depends on the properties of the primordial source of radiation. 

We employ the \texttt{21cmFAST} code \cite{Mesinger2011} to simulate the 21-cm signal measurement, $\delta T_\mathrm{b}$, between frequencies $200$ and $70\,\rm{MHz}$. We create a dataset with 15'945 samples by varying the cosmic initial conditions and six astrophysical parameters to obtain different reionization histories. The dataset is split into 12'000 samples for training (75.3$\%$), 2'000 for validation (12.5$\%$) and 1'945 for testing (12.2$\%$). These astrophysical parameters define the efficiency of the formation of luminous sources and the production rate of ionising photons; we treat them as nuisance parameters, namely, they do not constitute the main target of our inference process (see \cite{Park2019InferringSignal} for a detailed description).

Radio telescopes measure the 21-cm signal in Fourier space, proportional to the fluctuations in $\delta T_\mathrm{b}$, providing observations in terms of spatial frequency components. The primary observable from the initial SKA-Low datasets will be the 2D power spectrum, \( P(k_\perp, k_\parallel) \), where \( k_\perp \) and \( k_\parallel \) represent the transverse and line-of-sight wave numbers, respectively. To simulate this, we divide each realisation into three sub-volumes corresponding to frequency ranges \([151,\,165.9]\,\mathrm{MHz}\), \([166,\,180.9]\,\mathrm{MHz}\), and \([181,\,195.9]\,\mathrm{MHz}\). For each range, we compute \( P(k_\perp, k_\parallel) \) using the \texttt{tools21cm} package \cite{Giri2020t2c}. This quantity retains sensitivity to the underlying IGM ionization state: $P(k_\perp,k_\parallel) \propto \bar{x}_\mathrm{HI}^2$ \cite{Furlanetto2006CosmologyUniverse}, where $\bar{x}_\mathrm{HI}$ is the volume-averaged neutral fraction within the observed frequency range. 

\begin{figure}[t]
    \includegraphics[width=\textwidth]{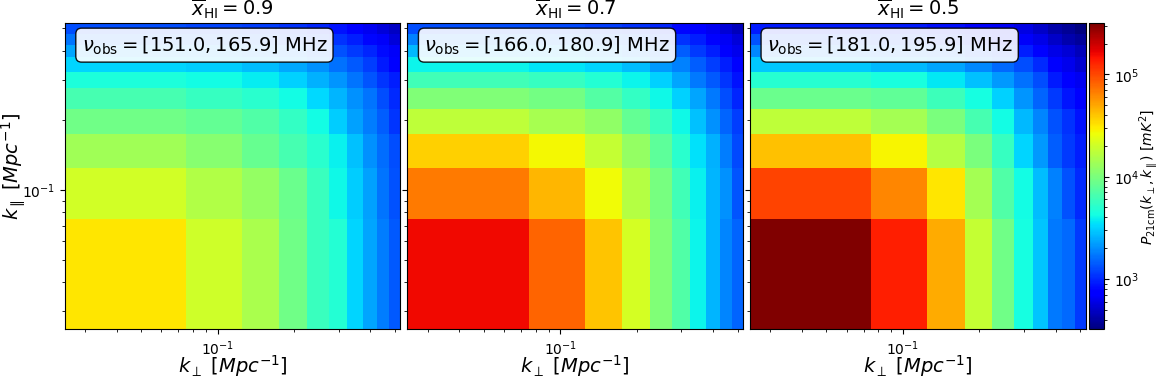}\vskip-4mm
    \caption{2D power spectra of the cosmological 21-cm signal measured at the three different observed frequency ranges for one model in our dataset. On top of each panel, we show the corresponding volume-averaged neutral fraction, $\overline{x}_\mathrm{HI}$.}
    \label{fig:pk_example}
\end{figure}

In \autoref{fig:pipeline}, we show an example of the inference pipeline for this paper. From each realisation of $\delta T_\mathrm{b}$ 3D SKAO mock observation data, we select three sub-volumes for the above frequency range and calculate the 2D power spectra, $P(k_\perp, k_\parallel)$. This 2D power spectra data is analysed to infer the EoR history ($\bar{x}_\mathrm{HI}$).
In \autoref{fig:pk_example}, we show the computed 2D power spectra of the model at the three observed frequency ranges. These $10\times10$ images constitute the input of our machine learning approaches, while the corresponding average neutral fraction, $\bar{x}_\mathrm{HI}$, at the observed frequency range is the target.

\subsection{Evaluation Methodology}
We employ two metrics to quantify the regression performed by the different deep learning methods. The first metric is the coefficient of determination, $R^2$, defined as:
\begin{equation}\label{eq:r2score}
    R^2 (y, \hat{y}) = 1 - \frac{\sum_i (y_i - \hat{y}_i)^2}{\sum_i (y_i - \bar{y})^2} \ .
\end{equation}
Here $\hat{y}$ is the prediction and $y$ is the ground truth, while $\bar{y} = \frac{1}{N}\sum_i y_i$ is the average over the test dataset at a given frequency range. The second metric is the root-mean-square error, $RMSE$, defined as:
\begin{equation}\label{eq:rmse}
    RMSE (y, \hat{y}) = \sqrt{\frac{1}{N}\sum_i (y_i - \hat{y}_i)^2} \ .
\end{equation}
In our case, $N$ is the number of samples for the test set. In \autoref{tab:results1} and \ref{tab:results2}, we compare the score on the test set for different deep learning methods. To ensure a fair comparison, all models presented in this paper are trained and evaluated on the same dataset, see \S\ref{sec:data}.

\subsection{Deep Learning Models} \label{sec:deeplr}
In this section, we present the models implemented to solve this challenge. We implemented and evaluated a broad selection of promising models highlighted by the literature and models that yielded high-performing results for similar tasks. 

\subsubsection{Generative Flow Network}
The GLOW (Generative Flow) architecture \cite{Kingma2018glow} builds upon one of the most widely employed architectures \cite{Kobyzev2021NFreview}, i.e., coupling flows. Each layer comprises three invertible transformations: an activation normalization (Act-Norm), a $1\times1$ invertible convolution, and an affine coupling operation.

Normalizing Flow (NF) networks learn a mapping between a complex data distribution $\hat{p}_Y$ and a simple base distribution $p_Z$ for the target and random variables, $\mathbf{Y},\,\mathbf{Z} \in \mathbb{R}^D$, respectively. A bijection function defines the mapping, $f: \mathbb{R}^D \to \mathbb{R}^D$, between the target random variable $\mathbf{Y} = f^{-1}(\mathbf{Z})$ and the random distribution. The mapping is composed of $N$ invertible transformations $f^{-1}(\mathbf{z})=f^{-1}_N \circ f^{-1}_{N-1} \circ \cdots \circ f^{-1}_{1}(\mathbf{z})$ referred as the coupling flow. We then consider a disjoint partition that splits the input in half $x^A,x^B\in\mathbb{R}^{D/2}$. The first part is processed by the coupling flow, $y^A=f^{-1}(x^A)$ while $x^B$ is processed by the $1\times1$ invertible convolution, $\Theta$, $y^B=f^{-1}(\Theta(x^B))$. The result is then concatenated and processed by the next layer. This approach gradually introduces dimension in the flow generative process, reducing computational cost while capturing the multi-scale structure of the high-dimensional distribution \cite{Dinh2017RealNVP}. 

The network is optimized by training and learning the parameters, $W\in\mathbb{R}^{D\times D}$, of the transformations $f$ such that the total likelihood of the observed data is maximized.

\subsubsection{SE-CNN}\label{subsec:secnn}
The SE-CNN architecture is a convolutional neural network augmented with Squeeze-and-Excitation (SE) blocks \cite{8578843}. Our proposed architecture consists of two convolutional layers with ReLU activation, each followed by batch normalization and max pooling. SE blocks are inserted after each convolution to re-weight channel-wise feature responses adaptively.

Each SE block performs global average pooling across spatial dimensions, followed by two fully connected layers with a bottleneck structure and a sigmoid activation to generate channel-wise weights. These are applied multiplicatively to the feature maps, allowing the network to modulate feature importance across channels.

This mechanism allows the model to dynamically emphasize more informative channels dynamically, enhancing its ability to capture relevant features while suppressing less useful ones. In their study on hyperspectral image classification, \cite{asker-2024} demonstrated that SE-based architectures improve classification performance by enabling more discriminative feature selection across spectral bands. Such channel-wise attention mechanisms can help reduce overfitting and improve generalization, particularly when the input features vary significantly in relevance. The convolutional backbone is followed by one or two dense layers, with dropout, and a final output layer for regression.

\subsubsection{SE-CNN Ensemble-10}
The SE-CNN Ensemble-10 consists of ten independently trained SE-CNN models, each initialized with a different random seed. The base architecture follows the design described in \S\ref{subsec:secnn} (SE-CNN), using SE blocks \cite{8578843} to adaptively re-weight channel-wise features after each convolutional layer.

The ensemble was implemented to enhance robustness and prediction stability. Each model was trained on the same dataset but converged to a different local minimum due to its unique initialization. At inference time, predictions from all models were averaged to produce the final output. This strategy is motivated by prior work showing that deep ensembles can effectively reduce variance and improve generalization in regression tasks \cite{lakshminarayanan-2016}.

\subsubsection{MLP-Mixer}
The MLP-Mixer is a neural network architecture that replaces convolution and attention mechanisms with MLPs for both spatial (token) and channel mixing. Our implementation adapts the design introduced in \cite{tolstikhin-2021}, using a series of fully connected layers applied over reshaped image patches.

\begin{figure}[]
    \centering
        \resizebox{0.9\textwidth}{!}{\begin{tikzpicture}[
  block/.style={rectangle, draw, minimum height=1cm, minimum width=2.9cm, text centered},
  tokenmlp/.style={rectangle, draw, fill=blue!15, minimum height=1cm, minimum width=3.2cm, text centered},
  channelmlp/.style={rectangle, draw, fill=orange!15, minimum height=1cm, minimum width=3.2cm, text centered},
  outputblock/.style={rectangle, draw, fill=purple!10, minimum height=1cm, minimum width=2.9cm, text centered},
  arrow/.style={thick, -{Stealth[]}},
  residual/.style={->, thick, dashed, bend left=30, color=green!60!black},
  repeatbox/.style={draw=black!60, dashed, thick, rounded corners},
  node distance=1.3cm and 1.3cm
  ]

\node[block] (input) {\shortstack{Input Image \\ $10 \times 10 \times 1$}};
\node[block, below=0.7cm of input] (reshape) {\shortstack{Reshape \& Flatten \\ $100 \times 1$}};
\node[block, below=0.7cm of reshape] (proj) {\shortstack{Linear Projection \\ $100 \times d$}};

\node[tokenmlp, right=of input,xshift=0.5cm] (token1) {Token-Mixing MLP};
\node[channelmlp, below=of token1] (channel1) {Channel-Mixing MLP};

\draw[residual] (token1.south east) to[out=45,in=200] node[midway, above] {\scriptsize Skip Conn.} (channel1.north west);

\node[tokenmlp, right=of token1] (token2) {Token-Mixing MLP};
\node[channelmlp, below=of token2] (channel2) {Channel-Mixing MLP};

\draw[residual] (channel1.north east) to[out=340,in=130] node[pos=0.64, below] {\scriptsize Skip Conn.} (token2.west);
\draw[residual] (token2.south east) to[out=45,in=200] node[midway, above] {\scriptsize Skip Conn.} (channel2.north west);

\node[outputblock, right=of token2, xshift=0.5cm] (pool) {Global Avg. Pooling};
\node[outputblock, below=0.7cm of pool] (dense) {Dense Layers};
\node[outputblock, below=0.7cm of dense] (output) {\shortstack{Output \\ Regression}};

\draw[arrow] (input) -- (reshape);
\draw[arrow] (reshape) -- (proj);
\draw[arrow] (proj) -- (token1);
\draw[arrow] (token1) -- (channel1);
\draw[arrow] (channel1) -- (token2);
\draw[arrow] (token2) -- (channel2);
\draw[arrow] (channel2) -- (pool);
\draw[arrow] (pool) -- (dense);
\draw[arrow] (dense) -- (output);

\begin{scope}[on background layer]
\node[repeatbox, fit=(token1)(channel1)(token2)(channel2), inner sep=10pt, label={[yshift=1mm]above:\textbf{Repeated Mixer Layers (×4)}}] {};
\end{scope}

\end{tikzpicture}}\vskip-3.4mm
        \caption{MLP-Mixer architecture adapted for 2D input, see \S\ref{sec:deeplr}. The model processes flattened image patches through repeated Mixer layers, each combining token-mixing and channel-mixing MLPs with skip connections. The final output is obtained via global average pooling and dense layers.} 
        \label{fig:mlp_mixer_final}
\end{figure}
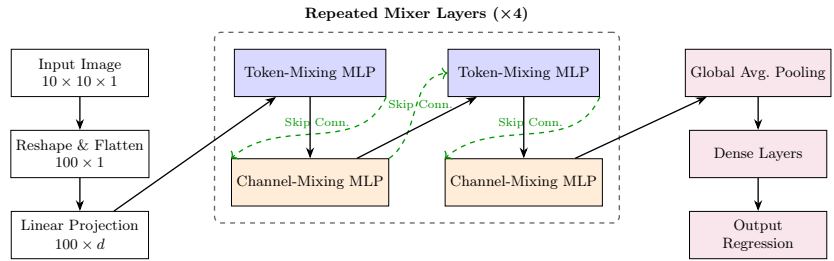

The model receives a $10\times10$ image reshaped into a sequence of 100 tokens, each treated as a spatial unit. Each token is linearly projected into a higher-dimensional space. A series of Mixer blocks is then applied, consisting of two stages: token-mixing and channel-mixing. In token-mixing, interactions across spatial positions are captured by transposing the token and channel axes, applying a shared MLP, and restoring the original shape. Each token is processed independently across its features in channel-mixing using another MLP.  In \autoref{fig:mlp_mixer_final}, we show a schematic representation of the architecture proposed.

Residual connections, layer normalization, and GELU \cite{hendrycks-2016} activations are used throughout. After the mixer layers, the output is globally averaged, passed through two fully connected layers with dropout, and finally mapped to a scalar output for regression.

\subsubsection{MiniViT}
MiniViT is a compact Vision Transformer (ViT) architecture tailored to the small input size of our cosmological data maps ($10 \times 10$ pixels). Our proposed architecture adapts the ViT framework \cite{dosovitskiy-2020} by simplifying the transformer depth and tokenization strategy to suit low-resolution inputs and reduce computational complexity.

The input image is first reshaped into a sequence of 100 tokens, each representing a pixel, and linearly projected into a higher-dimensional embedding space. A trainable positional encoding is added to each token embedding to retain spatial structure. The sequence is then processed through a stack of transformer encoder blocks. Each block consists of a multi-head self-attention mechanism and a feed-forward MLP with GELU activations wrapped in residual connections and layer normalization.

Following the transformer layers, the output embeddings are aggregated using global average pooling and passed through a dense regression head. The design preserves key transformer properties, such as global receptive fields and dynamic attention mechanisms, while maintaining computational efficiency. This is particularly important for low-dimensional inputs, where compact Vision Transformer variants have been shown to offer favorable trade-offs between performance and efficiency \cite{si-2025}. GELU activation layers provide a smooth and probabilistically motivated non-linearity, which has been shown to improve training dynamics in transformer architectures \cite{hendrycks-2016}.

\subsubsection{Frequency-Aware CNN}
We implemented a custom convolutional neural network architecture designed to condition on the frequency information observed. This design enables the model to incorporate auxiliary knowledge about the input's observational frequency band, which may influence the signal characteristics. The conditioning is achieved by injecting the frequency range as an additional input, spatially aligned to match the dimensions of the image input.

Specifically, the scalar frequency category is first one-hot encoded and reshaped into a small $1 \times 1 \times 3$ tensor. This tensor is then upsampled to the exact spatial resolution as the image (i.e., $10 \times 10$), effectively creating three additional constant feature maps. These are concatenated along the channel dimension with the original $10 \times 10 \times 1$ image, producing a combined input of shape $10 \times 10 \times 4$.

This approach allows the convolutional layers to access spatial and contextual frequency range information from the first layer, potentially improving the model's generalization ability across different spectral regimes. The idea of conditioning convolutional networks by concatenating auxiliary data to the input tensor has been effectively employed in other contexts, notably in conditional GANs \cite{mirza-2014}, and provides a simple yet powerful mechanism for context-aware learning. The rest of the architecture mirrors a typical CNN pipeline: two convolutional layers (each followed by batch normalization, ReLU activation, and max pooling), a flattening step, fully connected layers, and a regression output head.

\subsubsection{Simulation-Based Inference}
The physical processes underlying the EoR are inherently complex, and approximations like the Gaussian likelihood typically assumed in Bayesian analyses could significantly bias the final inference. SBI recently emerged as a principled framework to actually learn the likelihood (or analogous quantities following Bayes' theorem) from a set of fiducial simulations \cite{Cranmer20}, as those described in Sect.~\ref{sec:data}. We therefore develop an SBI pipeline to learn the posterior distribution $p (\mathbf{\theta}|\mathbf{d})$ from our set of simulations; in this case, the SBI task is usually dubbed neural posterior estimation (NPE). We employ the publicly available \texttt{sbi} package \cite{tejerosbi2022code}, which provides the infrastructure required to train a NF to learn the posterior distribution and apply it using the same data splits as in the previous sections. We consider two distinct SBI approaches: the \textit{marginal} prediction of each individual $\bar{x}_\mathrm{HI}$ (at different frequencies) together with the astrophysical parameters; and the \textit{joint} prediction of $\bar{x}_\mathrm{HI}$ at different frequencies but from the same simulation, ignoring the nuisance astrophysical parameters. In the latter case, the input of the NF consists of the stacked power spectra for each frequency. In principle, this provides more information to disentangle the effect of the simulation parameters from the EoR history.

\section{Results and Discussion}\label{sec:result}
\autoref{tab:results1} shows the overall performance of each implemented model on the full test dataset, while \autoref{tab:results2} shows the measured metric for three observed frequency ranges, see \S\ref{sec:data}. 

\begin{table}[]
    \centering
\caption{Performance comparison of different models on the test dataset.}
    \begin{tabular}{|l|M{2.0cm}|M{2.8cm}|}
        \hline
        \textbf{Model} & \boldmath$R^2$ [\%] $\uparrow$ & \textbf{RMSE} $\downarrow$\\
        \hline\hline
        GLOW & $98.09  $ & $3.72\times 10^{-2}$ \\
        SBI (marginal) & 88.04  & $9.31 \times 10^{-2}$ \\
        SBI (joint) & 97.44 &  $4.23 \times 10^{-2}$ \\
        SE-CNN & 98.06  & $3.75 \times 10^{-2}$ \\
        SE-CNN Ens.-10 & $\mathbf{98.61}$ & $\mathbf{3.18 \times 10^{-2}}$ \\
        MLP-Mixer & 98.58  & $3.21 \times 10^{-2}$ \\
        MiniViT & 95.55 & $5.67 \times 10^{-2}$ \\
        Freq.-Aware CNN & 98.43 & $3.37 \times 10^{-2}$  \\
        \hline
    \end{tabular}
    \label{tab:results1}
\end{table}

\begin{table}[]
    \caption{Summary of the metrics on the test dataset for the different methods, split by frequency range.} \label{tab:score}
    \begin{tabular}{|l|M{1.5cm}|M{1.75cm}|M{1.5cm}|M{1.75cm}|M{1.5cm}|M{1.75cm}|} \cmidrule{2-7}
     \multicolumn{1}{c|}{} & \multicolumn{2}{c|}{$[151,166]\,\mathrm{MHz}$} & \multicolumn{2}{c|}{$[166,181]\,\mathrm{MHz}$} & \multicolumn{2}{c|}{$[181,196]\,\mathrm{MHz}$} \\ \cmidrule{1-7}
     \textbf{Model} & \boldmath$R^2$ [$\%$] $\uparrow $& \textbf{RMSE} $\downarrow $ & \boldmath$R^2$  [$\%$] $\uparrow $& \textbf{RMSE} $\downarrow$& \boldmath$R^2$ [$\%$] $\uparrow$ & \textbf{RMSE} $\downarrow$ \\ \hline\hline
     GLOW & $95.76$ & $3.87\times10^{-2}$ & $97.75$ & $3.65\times10^{-2}$ & $98.41$ & $3.62\times10^{-2}$ \\ 
     SBI (marginal)  &   88.08  & $6.50\times10^{-2}$ &  78.03     & $11.42\times10^{-2}$ & 89.40 &  $9.37\times10^{-2}$  \\ 
     SBI (joint)  & 94.50  & $4.17\times10^{-2}$  & 96.53 &  $4.40\times10^{-2}$ & 97.93 &  $4.10\times10^{-2}$ \\ 
     SE-CNN & $97.69$ & $2.84\times10^{-2}$ & $97.84$ & $3.57\times10^{-2}$ & $97.98$ & $4.08\times10^{-2}$ \\
     SE-CNN Ens.-10 & $97.96$ & $2.68\times10^{-2}$ & $98.10$ & $3.36\times10^{-2}$ & $\mathbf{98.47}$ & $\mathbf{3.56\times10^{-2}}$ \\
     MLP-Mixer & $\mathbf{98.41}$ & $\mathbf{2.37\times10^{-2}}$ & $\mathbf{98.30}$ & $\mathbf{3.17\times10^{-2}}$ & $98.25$ & $3.80\times10^{-2}$ \\
     MiniViT & $92.94$ & $4.97\times10^{-2}$ & $94.37$ & $5.77\times10^{-2}$ & $94.31$ & $6.82\times10^{-2}$ \\
     Freq.-Aware CNN & $98.11$ & $2.55\times10^{-2}$ & $97.82$ & $3.59\times10^{-2}$ & $97.98$ & $4.07\times10^{-2}$ \\ \hline

    \end{tabular}\label{tab:results2}
\end{table}


Our benchmark study across multiple deep learning architectures reveals consistently high performance in predicting the neutral hydrogen fraction from 2D 21-cm power spectra. Among the models, the SE-CNN Ensemble-10 achieves the highest overall performance on the test set, benefiting from the variance reduction and robustness typically provided by deep ensembles. However, when evaluating performance across individual frequency ranges, the MLP-Mixer slightly outperforms the ensemble in two out of three bands and shows remarkably stable results throughout. Despite being the second-best model in terms of global metrics ($R^2 = 98.58\%$), its consistency across observational conditions highlights its strong generalization capabilities. This divergence between aggregate and group-wise results is reminiscent of Simpson's paradox \cite{simpson1951interaction}, where trends observed in subgroups can be masked when data is pooled. Together, these results suggest that the MLP-Mixer is an exceptionally reliable architecture under varying data regimes and may benefit further from ensemble strategies.

Notably, the Frequency-Aware CNN, a custom model explicitly conditioned on the frequency band via one-hot encoded inputs, performs nearly on par with ensemble and attention-based models. This shows that integrating frequency context can be just as effective as channel attention mechanisms like SE blocks.

By contrast, the MiniViT architecture underperforms, with $R^2$ scores consistently below 96\%. During training, this model exhibited slow convergence and high variance, likely reflecting the known data inefficiency of transformer-based models, which generally require large-scale datasets and extensive pretraining to reach optimal performance \cite{dosovitskiy-2020,touvron2021trainingdataefficientimagetransformers}. This underscores a key limitation of applying ViT-style models directly on small cosmological datasets without tailored adaptations.

The GLOW architecture shows an increasing accuracy for increasing frequency, starting from the low frequency range at $R^2\approx 95 \%$ and $RMSE\simeq3.8\times10^{-2}$ up/down to $R^2\approx 98 \%$ and $RMSE\simeq3.6\times10^{-2}$. This trend follows the signal evolution in the input data (the 2D power spectra) as shown in \autoref{fig:pk_example}, indicating that the network is sensitive to the fluctuations of the 21-cm signal. If not accounted for, we expect instrumental noise to decrease the accuracy of the network, as systematics will increase the signal-to-noise ratio and break the signal evolution.

Regarding SBI, the joint model outperforms the marginal approach. This is the consequence of more information being provided to the network and demonstrates the importance of including all frequencies together to break degeneracies between $\bar{x}_\mathrm{HI}$ and the astrophysical parameters of the simulations. It is noteworthy that the joint model performs nearly on par with several CNN-based architectures. In contrast, the SBI model does not take advantage of the 2D nature of the data, since the input is flattened.

\section{Conclusion}
In this paper, we implemented a broad selection of various Deep Learning models in the hope of progressing the 21-cm signal extraction, a complex task that traditional approaches struggle with. Our models were tested on a dataset we generated according to the SKA specifications.
Several models performed quite well, especially ensemble CNN method and MLP-Mixer, with a maximum $R^2$ score of 98.61\%. Our approaches could be used and tested on real data when the SKAO will be operational in the coming years.

\begin{credits}
\subsubsection{\ackname} The authors acknowledge access to Piz Daint at the Swiss National Supercomputing Centre, Switzerland, under the SKA share with the project ID sk014. The authors acknowledge funding from the Spark grant CRSK-2\_228671 from the Swiss National Science Foundation, as well as support from the Sweden SKA Regional Center (sweSRC) node operated by Onsala Space Observatory in collaboration with Chalmers e-Commons. The Onsala Space Observatory national research infrastructure is funded through Swedish Research Council grant No 2019-00208.

\subsubsection{\discintname} The authors have no competing interests to declare that are relevant to the content of this article.

\end{credits}
%
%
%
\bibliographystyle{splncs04}
\bibliography{bibliography}
\end{document}